\documentclass[twocolumn]{aastex701}
\usepackage{amsmath}

\newcommand{\aperture}{\emph{Aperture}}

\definecolor{ao(english)}{rgb}{0.0, 0.5, 0.0}

\begin{document}

\title{Stable Collisionless Tori Around Kerr Black Holes}

\author[0009-0004-3333-7897]{Martin Luepker}
\affiliation{Physics Department and McDonnell Center for the Space Sciences, Washington University in St. Louis, MO 63130, USA}
\email[show]{luepker.m@wustl.edu} 

\author[0000-0002-0108-4774]{Yajie Yuan}
\affiliation{Physics Department and McDonnell Center for the Space Sciences, Washington University in St. Louis, MO 63130, USA}
\email[]{yajiey@wustl.edu} 

\author[0000-0002-4738-1168]{Alexander Y. Chen}
\affiliation{Physics Department and McDonnell Center for the Space Sciences, Washington University in St. Louis, MO 63130, USA}
\email[]{cyuran@wustl.edu}

%% Use the \collaboration command to identify collaborations. This command
%% takes an optional argument that is either a number or the word "all"
%% which tells the compiler how many of the authors above the command to
%% show. For example "\collaboration[all]{(DELVE Collaboration)}" wil include
%% all the authors above this command.
%%
%% Mark off the abstract in the ``abstract'' environment. 
\begin{abstract}
In low-luminosity active galactic nuclei like M87$^\ast$ and Sgr A$^\ast$, the accretion flow in the vicinity of the black hole is in the collisionless regime, meaning that the collisional mean free path of charged particles is much larger than the dynamical length scales. To properly model the particle energization and emission from the collisionless accretion flow, a promising approach is to employ the global general-relativistic particle-in-cell simulations---a newly developed, fully kinetic, first-principles method. However, it has been challenging to set up an initial condition that involves collisionless gas with finite angular momentum. We present, for the first time, a class of analytic kinetic equilibria of collisionless tori around a Kerr black hole. We have successfully implemented the collisionless tori in our GPU-based GRPIC code framework \aperture{}, and found them to be stable for hundreds to thousands of dynamical times in 2D axisymmetric simulations when there is no initial seed magnetic field. These kinetic equilibria serve as ideal starting points for future studies of the physics of collisionless accretion and jet launching. 
\end{abstract}

%% Keywords should appear after the \end{abstract} command. 
%% The AAS Journals now uses Unified Astronomy Thesaurus (UAT) concepts:
%% https://astrothesaurus.org
%% You will be asked to selected these concepts during the submission process
%% but this old "keyword" functionality is maintained in case authors want
%% to include these concepts in their preprints.
%%
%% You can use the \uat command to link your UAT concepts back its source.
\keywords{\uat{Black hole physics}{159} --- \uat{High energy astrophysics}{739} --- \uat{Plasma astrophysics}{1261}}

%% From the front matter, we move on to the body of the paper.
%% Sections are demarcated by \section and \subsection, respectively.
%% Observe the use of the LaTeX \label
%% command after the \subsection to give a symbolic KEY to the
%% subsection for cross-referencing in a \ref command.
%% You can use LaTeX's \ref and \label commands to keep track of
%% cross-references to sections, equations, tables, and figures.
%% That way, if you change the order of any elements, LaTeX will
%% automatically renumber them.

\section{Introduction}
\label{sec:intro}
Accretion onto supermassive black holes powers the fascinating phenomena of active galactic nuclei (AGN). As the gas accretes toward the black hole, its gravitational potential energy is converted into heat and radiation, making the nucleus of the galaxy shine across a wide range of the electromagnetic spectrum. Some of the accreting black holes can also launch powerful, relativistic jets extending well beyond the galaxy itself, which, in turn, shape the environment of the galaxy.

Recently, the Event Horizon Telescope (EHT) has made unprecedented, high-resolution observations of nearby supermassive black holes, allowing us to directly see the central engines in action. In particular, the EHT obtained polarized images of M87$^\ast$ and Sgr A$^\ast$~\citep{2021ApJ...910L..12E,2021ApJ...910L..13E,2024ApJ...964L..25E,2024ApJ...964L..26E}, showing that the bright ring surrounding the black hole shadow is the radiation from the plasma accreting toward the black hole, and that there is a strong magnetic field near the horizon, which may be the key for jet launching. Currently, the best models for the horizon-scale emission come from general relativistic magnetohydrodynamics (GRMHD) simulations~\citep{2019ApJ...875L...5E,2022ApJ...930L..16E}. However, GRMHD cannot self-consistently capture the energization of the emitting electrons, leading to significant model uncertainties. Additionally, no model in the GRMHD library has been found to satisfactorily reproduce all the observed features of Sgr A$^\ast$. These models cannot simultaneously explain the variability and geometric properties of this source~\citep{2022ApJ...930L..16E}, suggesting that they may be missing some important physical ingredients.

M87$^\ast$ and Sgr A$^\ast$ belong to a class of AGN with low-luminosity. In these systems, the gas density near the black hole is so low that the plasma is collisionless---the mean free path of Coulomb collisions between charged particles is larger than the system size. Take M87$^\ast$ as an example, a simple one-zone model based on EHT observations gives an estimate of the electron density in the emitting region $n_e\sim 10^{4\text{--}7}\,{\rm cm}^{-3}$ and electron temperature $T_e\sim(1\text{--}12)\times 10^{10}\,{\rm K}$~\citep{2021ApJ...910L..13E}.
We can then estimate the mean free path for electrons colliding with ions as $\lambda_{\rm mfp}\sim (kT_e)^2/(\pi n_e e^4)\sim 10^{19}T_{e,10}^2n_{e,6}^{-1}\,{\rm cm}$ (we use the notation $T_{e,10}\equiv T_e/(10^{10}\,{\rm K})$, $n_{e,6}\equiv n_e/(10^6\,{\rm cm}^{-3})$).  
Clearly, $\lambda_{\rm mfp}$ is much larger than the gravitational radius $r_g\equiv GM/c^2\approx 9.6\times 10^{14}\,{\rm cm}$, where we have used a mass of $M=6.5\times10^9M_{\odot}$ for M87$^\ast$~\citep{2019ApJ...875L...1E}. Therefore, the plasma is collisionless. MHD typically assumes that the plasma behaves like an ideal gas that locally reaches thermal equilibrium, but this requires efficient collisions. In the collisionless regime, plasmas do not necessarily maintain local thermal equilibrium; electrons and ions can have quite different distributions. Kinetic effects are important to determine the thermodynamics of the plasma, especially differential heating of electrons and ions, and non-thermal particle acceleration, which cannot be captured in MHD simulations.

Recently, significant progress has been made in developing a first-principles, kinetic simulation method---general relativistic particle-in-cell (GRPIC)---to study the collisionless plasma dynamics around black holes~\citep[e.g.,][]{2019PhRvL.122c5101P,2020PhRvL.124n5101C,2021A&A...650A.163C,2022PhRvL.129t5101C,2021PhRvL.127e5101B,2022A&A...663A.169E,2023A&A...677A..67E,2025arXiv250401062M,2025arXiv250304558C,2025ApJ...985..159Y}. The GRPIC approach follows the evolution of the electromagnetic field and particle distribution function self-consistently by solving the coupled Maxwell and Vlasov equations in the curved spacetime. The approach retains the full information of collective plasma behavior as well as nonthermal particle acceleration. Global GRPIC simulations have just started to be employed to study the collisionless accretion process \citep{2023PhRvL.130k5201G,2025PhRvL.135a5201V}, which demonstrate different behaviors compared to MHD, particularly with respect to magnetic reconnection, instabilities caused by plasma anisotropy, heat flux, and plasma mixing at the jet/disk boundary. However, the studies so far have only considered accretion of zero-angular-momentum gas, which is an oversimplification---in reality, the gas reservoir at the center of galaxies likely has nonzero angular momentum. To understand the accretion physics in low-luminosity AGN, we need to focus on more realistic configurations.

We would like to build a self-consistent model of collisionless accretion flows, starting from a physical initial condition where the gas supply near the black hole has finite angular momentum. A commonly used initial condition in GRMHD simulations is a hydrodynamic equilibrium assuming a perfect fluid~\citep[e.g.,][]{1976ApJ...207..962F,1978A&A....63..221A}.
However, initializing such a torus in the collisionless regime already presents the first challenge. For the perfect fluid torus, the equilibrium is possible because a combination of the gravitational force and the pressure gradient provide the centripetal force for the orbiting gas element. 
The perfect fluid assumption requires that the stress-energy tensor takes the simple form $\mathrm{diag}(\varepsilon, P,P,P)$ in the fluid rest frame, where $\varepsilon$ is the total energy density and $P$ is the pressure.

In the collisionless regime, 
if we set up a plasma distribution that is locally Maxwellian
following an instance of the Fishbone-Moncrief torus~\citep[e.g.,][]{2003ApJ...589..444G},
it turns out that the 
distribution quickly evolves away from the initial configuration, even when electromagnetism is turned off. 
This is because particles are freely streaming along geodesics and there is no way to maintain 
the perfect fluid assumption as required by these hydrodynamic torus solutions.

In this paper, we present, for the first time, a class of analytic kinetic equilibria of collisionless tori around a Kerr black hole. We have successfully implemented the collisionless tori in our GPU-based GRPIC code framework \aperture{}, and found them to be stable for hundreds to thousands of dynamical times in 2D axisymmetric simulations when there is no initial seed magnetic field. These kinetic equilibria serve as ideal testbeds for us to further study the physics of collisionless accretion and jet launching.

\begin{figure*}[ht!]
\plotone{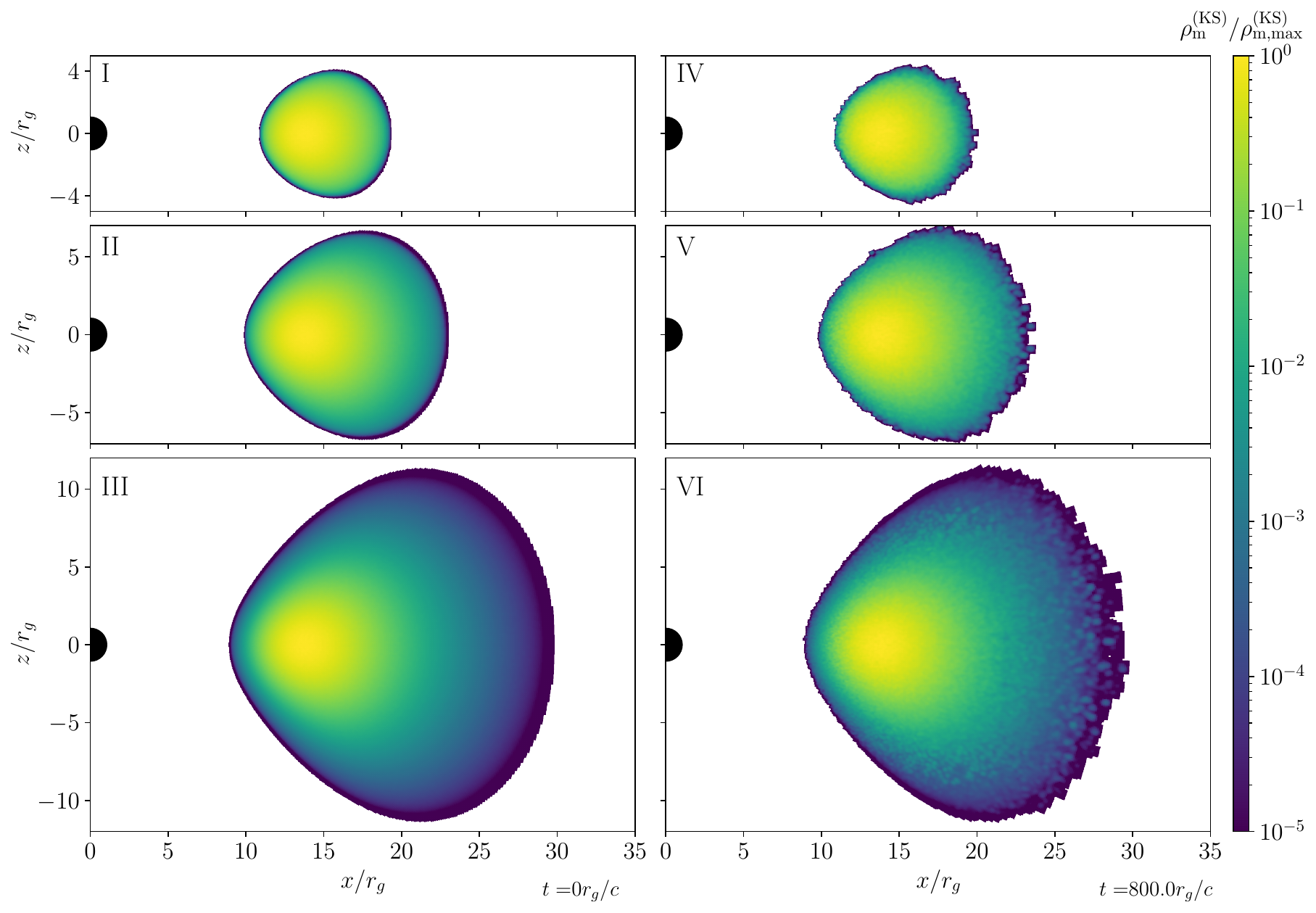}
\caption{ 
Cross sections of the density distribution for three axisymmetric tori. Panels~I, II, and III correspond to the initial tori with inner radii $r_{\mathrm{in}} = 11\,r_g$, $10\,r_g$, and $9\,r_g$, respectively, each normalized to unity. Panels~IV, V, and VI show these same tori after evolving for $800\,r_g/c$. We include electromagnetic effects but there is no initial background field. All cases have temperature $T=10^{-3}m$ and constant angular momentum $L_0 \approx 3.986\,m$, corresponding to $r_0=14\,r_g$ around a black hole with spin $a = 0.999$. The fluctuations seen in the right column are consistent with particle shot noise for 20 particles per cell. In each panel, the black circle shows the black hole event horizon, $r_h = r_g(1 + \sqrt{1 - a^2})$.}

\label{fig:density}
\end{figure*}

\section{Analytic collisionless tori}\label{sec:analytic}

\subsection{Vlasov Equation and Jeans's Theorem}\label{subsec:jeans}

To describe a collisionless ensemble of particles in general relativity, we use a distribution function $f(x^\mu,p_\mu)$ which gives the density of particles in phase space~\citep{1989ApJ...344..146R}. Here $x^{\mu}=(x^0,x^1,x^2,x^3)$ is the position 4-vector, $x^0=t$ is the coordinate time, and $p_{\mu}=(p_0,p_1,p_2,p_3)$ is the covariant 4-momentum. We use geometrized units such that $c=G=k_B=1$; Greek indices take on values $0,1,2,3$ while Latin indices take on values $1,2,3$. Throughout this paper, we use three different coordinate systems: Boyer--Lindquist (BL), Kerr--Schild (KS) and fluid rest frame (FRF). BL coordinates are most convenient for the analytic calculations, KS coordinates are used in the simulation because there is no coordinate singularity at the horizon and the FRF is required when calculating any local properties of the gas. Now, the number of particles $dN$ in a phase space volume element $dV_x dV_p$ is given by 
\begin{align}
    f=\frac{dN}{dV_x dV_p}.
\end{align}
In coordinate form, the invariant spatial volume element and the invariant momentum space volume are 
\begin{align}
    dV_x=\frac{p^0}{m}\sqrt{-g}\,dx^1dx^2dx^3, %d^3x 
    &&
    dV_p=\frac{m\, dp_1dp_2dp_3}{p^0\sqrt{-g}},
\end{align}
respectively, where $g$ is the metric determinant. The basics of the Kerr metric and the conventions used in this paper are listed in Appendices~\ref{sec:metric} and~\ref{sec:splitting}.

We consider the case where there is no electromagnetic field, and particles move freely in the Kerr spacetime following geodesics. Particle motion can be described using the Hamiltonian $\mathcal{H}=g^{\mu\nu}p_\mu p_\nu/2=-m^2/2$, which gives Hamilton's equations
\begin{align} \label{equ:hamilton}
    \frac{d x^\mu}{d\lambda}=\frac{\partial \mathcal{H}}{\partial p_\mu}, 
    && \frac{d p_\mu}{d\lambda}=-\frac{\partial \mathcal{H}}{\partial x^\mu},
\end{align}
where $\lambda$ is an affine parameter and the proper time is related to it by $\tau=m\lambda$.
According to Liouville's theorem, the distribution function $f$ is conserved along geodesics:
\begin{equation}\label{eq:Vlasov}
    \frac{df}{d\lambda}=0,
\end{equation}
This is the collisionless Boltzmann (Vlasov) equation. In coordinate form, Equation \eqref{eq:Vlasov} can be written as
\begin{equation}\label{eq:Vlasov_split}
    \frac{\partial f}{\partial t}+\left(\frac{dx^j}{dt}\right)\frac{\partial f}{\partial x^j}+\left(\frac{dp_j}{dt}\right)\frac{\partial f}{\partial p_j}=0,
\end{equation}
with $f=f(t,x^j,p_j)$. 
A steady state distribution needs to satisfy Equation \eqref{eq:Vlasov_split} with $\partial f/\partial t=0$, or equivalently, expressed using the Poisson bracket~\citep[e.g.][]{2023PhRvD.108l4022L}

\begin{align}\label{equ:boltzmann_Poisson_bracket}
       \{\mathcal{H},f\} =0.
\end{align}
We can construct steady state solutions utilizing Jeans's theorem \citep{Jeans_1929,thorne_modern_2017}: if $f$ depends on location $(x^j,p_j)$ in phase space only through constants of motion, then it is automatically a solution of the time independent Vlasov equation~\eqref{equ:boltzmann_Poisson_bracket}.

In Kerr spacetime, test particle trajectories have four constants of motion. With the symmetries of the Kerr metric, Hamilton's equations \eqref{equ:hamilton} provide two constants of motion,
energy $E=-p_0\ $and angular momentum $L_z=p_\phi$. The third constant of motion is the particle's rest mass $m=(-p^\mu p_\mu)^{1/2}$. \cite{1968PhRv..174.1559C} showed that there is a fourth constant $Q$ now called Carter's constant derived in Appendix \ref{sec:separation}. This allows us to write the Boyer--Lindquist momentum $p_\mu$ as a function of these constants and $r,\theta$ 
\begin{align}\label{equ:momentum}
    p_0=-E, &&
    p_r=\pm\frac{\sqrt{R}}{\Delta}, &&
    p_\theta=\pm\sqrt{\Theta},  &&
    p_\phi=L_z,
\end{align}
where $R$ and $\Theta$ are given in \eqref{eq:radial_function} and \eqref{eq:polar_function}.
We can then construct steady state distribution functions by writing $f(x^j,p_j)=\Psi(E,L_z,Q,m)$, where $\Psi$ is some arbitrary functional form. In what follows, we 
propose a simple
physically motivated functional form $\Psi$ such that the steady state is a bounded torus around the Kerr black hole.

\subsection{Collisionless Torus Distribution} \label{subsec:model}
To write down the analytic solutions, we work in the Boyer--Lindquist coordinates and use the $3+1$ formalism ~\citep[][and Appendix \ref{sec:splitting}]{2004MNRAS.350..427K}. It is well known that, at constant angular momentum, there is a range of energies that correspond to stable geodesic orbits ~\citep[e.g.,][]{1972ApJ...178..347B, 2008PhRvD..77j3005L}. 
We choose a distribution function with constant angular momentum $L_0$ and maximum energy $E_{\max}$ that confines the geodesics to a finite region. We choose a Maxwell-Boltzmann distribution in $E$ with exponential scale $T$ ~\citep[as in][]{2017JPhCS.831a2009R,2023PhRvD.108l4022L}
\begin{align}\label{equ:distribution}
  \begin{split}
    f(x^j,p_j) &= f_0\exp\left(\frac{E_{\max}-E}{T}\right) \\
     &\quad \times H(E_{\max}-E)\delta(L_z-L_0),
  \end{split}
\end{align}
where $H$ is the Heaviside function, and $f_0$ is a normalization constant. We call $T$ our ``energy temperature'', which should not be identified with the temperature of the plasma measured in its rest frame~(see Section~\ref{sec:fluid}).
This distribution describes an equilibrium torus that automatically satisfies \eqref{equ:boltzmann_Poisson_bracket} by Jeans's theorem. 
The energy $E$ in terms of the spatial and momentum coordinates $(x^j,p_j)$ is
\begin{align}\label{equ:energy}
    E= \alpha \sqrt{\gamma^{ij}p_i p_j+m^2}-\beta^i p_i,
\end{align}
where $\alpha$ is the lapse function, $\beta^i$ is the shift vector, and $\gamma^{ij}$ is the induced metric in the 3-space (Appendix \ref{sec:splitting}).
Note that within the torus, particle energy has to be greater than a lower bound $E_{\min}(r,\theta)$ in order to reach a certain location $(r,\theta)$. This minimum energy corresponds to a trajectory with $p_r=0$ and $p_\theta=0$ at that location, therefore
\begin{align}
    E_{\min}(r,\theta)=\alpha \sqrt{\gamma^{\phi\phi}L_0^2+m^2}-\beta^\phi L_0.
\end{align}
$E_{\min}(r,\theta)$ is position dependent. The global minimum, denoted as $E_0$, is located on the equator where the particle trajectory is a Keplerian circular orbit. Denote this radius as $r_0$, then we have~\citep{1972ApJ...178..347B}
\begin{align}
    L_0&=m\frac{a^2M^{1/2}r_0^{-3/2}-2aMr_0^{-1}+\sqrt{r_0M}}{\sqrt{1-3M r_0^{-1}+2a M^{1/2}r_0^{-3/2}}},\label{eq:circular_amom}\\
    E_0&=m\frac{1-2Mr_0^{-1}+aM^{1/2}r_0^{-3/2}}{\sqrt{1-3M r_0^{-1}+2a M^{1/2}r_0^{-3/2}}}\label{eq:circular_energy}.
\end{align}
Away from the point $(r_0,\pi/2)$, $E_{\min}(r,\theta)$ increases toward the edge of the torus, reaching $E_{\max}$ exactly at the boundary.
In practice, we determine a torus solution by first selecting $r_0$, which gives $L_0$ and the global minimum energy $E_0$. We then choose the inner radius of the torus on the equator $r_{\rm in}$, which determines the maximum energy $E_{\max}=E_{\min}(r_{\rm in},\pi/2)$. 
With these two parameters, $L_0$ and $E_{\max}$, the shape of the torus is completely defined, including its outer radius $r_{\rm out}$. The last parameter, energy temperature $T$, controls how quickly the density of the torus drops towards the boundary. The three parameters together define a series of equilibrium tori, in the same spirit as~\citet{1976ApJ...207..962F}.

\section{Implementation in GRPIC simulations}\label{sec:implementation}

We now describe how the analytic equilibrium torus described in Section~\ref{sec:analytic} can be implemented in GRPIC simulations. To solve the coupled Vlasov-Maxwell system, a source term that corresponds to the 4-current density $J^\mu$ needs to be computed from the particle distribution:
\begin{equation}
    J^\mu = \sum_s q_s\int p^\mu f \frac{d^3 p}{p^0\sqrt{-g}} = \sum_s q_s S^\mu_s,
\end{equation}
where $S^\mu$ is the number-flux 4-vector, and the sum is taken over all particle species. In particular, the charge density $J^0$ is proportional to the 0th component of $S^\mu$, and it will determine the particle density in the simulation. In addition, the local particle momentum distribution needs to be sampled correctly, which is often a nontrivial task for PIC simulations with nonstandard initial distribution functions.
In what follows, we describe the construction of the number-flux four-vector that determines the particle density and the sampling scheme for particle momenta.
\subsection{Change of Variables}\label{sec:changeofvariables}

To facilitate integration over momentum space and enable efficient sampling of particle velocities, we introduce polar coordinates in momentum space (in Boyer--Lindquist coordinates):
\begin{align}\label{equ:transformation}
    p_r=\frac{C}{\sqrt{\gamma^{rr}}}\hat p \cos\psi &&
    p_\theta=\frac{C}{\sqrt{\gamma^{\theta\theta}}}\hat p\sin\psi
\end{align}
where $C=\sqrt{\gamma^{\phi\phi}L_{0}^2+m^2}$, $\hat p\in [0,\hat p_{\max}]$ and $\psi\in [0,2\pi)$ such that $E(\hat p_{\max})=E_{\max}$. When making this change of variables to the distribution, the volume element picks up a Jacobian factor 
$\left|d^2p/(d\hat{p}d\psi)\right|$, 
thus
\begin{align}
    \delta(p_\phi-L_0 )d^3p=\frac{C^2\hat p \,d\hat{p}\,d\psi}{\sqrt{\gamma^{rr}\gamma^{\theta\theta}}}.
\end{align}

\subsection{Number-Flux Vector and Density Profile}

In order to compute the number-flux 4-vector $S^\mu$, it
is convenient to first compute the covariant components $S_\mu$, which have simpler dependence on the constants of motion, then raise the indices using the inverse metric. Substituting the change of variables above, we find
\begin{align}
  \begin{split}
    S_\mu &=\frac{f_0\alpha C \exp\left(\displaystyle\frac{E_{\max}+\beta^{\phi}L_0}{T}\right)}{\sqrt{\Delta \sin^2\theta}} \\
    &\quad \times \int \frac{p_\mu \hat p \exp\left(-\frac{\sqrt{\hat p^2+1} }{\hat{T}}\right)}{\sqrt{\hat p^2+1}} d\hat{p}\,d\psi
  \end{split}
\end{align}
where $\hat{T}=T/(\alpha C)$.
It is apparent that  $S_r=S_\theta=0$ since the integral of both $\sin\psi$ and $\cos\psi$ over $\psi$ are zero. The other two components can be evaluated analytically by substituting \eqref{equ:energy} or $p_{\phi}=L_z=L_0$, resulting in 
\begin{align}
    S_0&=\left.\frac{2\pi f_0  T(E+T)}{\sqrt{\Delta \sin^2\theta}}\exp\left(\frac{E_{\max}-E}{T}\right)\right|^{E_{\max}}_{E_{\min}},\\
    S_\phi &=-\left.\frac{2\pi f_0 TL_0}{ \sqrt{\Delta \sin^2\theta}}\exp\left(\frac{E_{\max}-E}{T}\right)\right|^{E_{\max}}_{E_{\min}}.
\end{align}

The expression for $S^0$ in terms of conserved quantities is obtained by raising the index using $g^{0\mu}$:
\begin{align}\label{equ:S^0}
    S^0=\left.\frac{2\pi f_0 T(T+E+\beta^\phi L_0)}{\alpha^2\sqrt{\Delta \sin^2\theta}}\exp\left(\frac{E_{\max}-E}{T}\right)\right|^{E_{\min}}_{E_{\max}}.
\end{align}
In a GRPIC code using the $3+1$ formalism~\citep{2004MNRAS.350..427K}, it is often convenient to use the convention that charge density is $\rho = \alpha J^0$, where $\alpha$ is the lapse function of the coordinate system used. This is the charge density measured by the fiducial observer (FIDO) of the coordinate system. We use the same convention in this paper. For each species $s$, the mass and charge densities are defined as
\begin{equation}
    \rho_{m,s} \equiv m_s\alpha S^0_s,\quad \rho_s \equiv q_s \alpha S^0_s.
\end{equation}
To ensure that the torus has no net charge or current, we assume that the negative charges and positive charges follow the same distribution function \eqref{equ:distribution} with the normalization such that $q_-f_{0,-}+q_+f_{0,+}=0$.

\subsection{Sampling Particle Momenta}
\label{subsec:momentum_sampling}

To sample particle momenta from the equilibrium distribution, we perform inverse transform sampling based on the probability density function after the coordinate transformation \eqref{equ:transformation}
\begin{align}\label{simplifiedMomentumDistribution}
    f(\hat p)&\propto \frac{\hat{p}}{\sqrt{\hat{p}^2+1}}\exp\left(-\frac{\sqrt{\hat p^2+1} }{\hat {T}}\right),
\end{align}
where all prefactors are omitted since they only depend on position which are normalized away when sampling. The cumulative distribution function (CDF) is:
\begin{align}
    F(\hat p)&=\frac{1}{Z}\int_{0}^{\hat p}\frac{ \hat p\exp\left(-\frac{ \sqrt{\hat p^2+1}}{\hat{T}}\right)}{\sqrt{\hat p^2+1}}d\hat p\\
    &=\frac{\hat{T}}{ Z}\left(\exp\left(-\frac{1 }{\hat{T}}\right)-\exp\left(-\frac{ \sqrt{\hat p^2+1}}{\hat{T}}\right)\right),
\end{align}
where $Z$ is the normalization constant ensuring $F(\hat{p}_{\max}) = 1$. This CDF can be readily inverted:
\begin{align}
    F^{-1}(u)=\sqrt{\left(\hat{T}\ln\left(\exp\left({-\frac{1}{\hat{T}}}\right)-\frac{uZ }{\hat{T}}\right)\right)^2-1},
\end{align}
where $u \in [0,1]$ is a uniform random variable. The angle $\psi$ is sampled uniformly in $[0,2\pi)$, and the full momentum covector is reconstructed using Equation~\eqref{equ:transformation}. 
\section{Simulations}\label{sec:simulation}

We use our GPU-based GRPIC code framework \aperture{}\footnote{The full code, including the implementation of the torus setup introduced in this paper, is hosted at \href{https://github.com/fizban007/Aperture4}{https://github.com/fizban007/Aperture4}.}~\citep{2025arXiv250304558C} to follow the evolution of the analytic, collisionless torus initial condition. For simplicity, we consider an electron-positron pair plasma.
The simulations are carried out in 2D axisymmetry using Kerr--Schild spherical coordinates, adopting the $3+1$ formalism of \citet{2004MNRAS.350..427K}. The benefit of this coordinate system is that there is no singularity at the event horizon, unlike the Boyer--Lindquist coordinates, and we can choose the domain boundary to be within the horizon. Similar choices are made in other GRPIC codes such as \citet{2019PhRvL.122c5101P,2025arXiv250708942M}. However, the formalism developed in Section~\ref{sec:implementation} is based on the Boyer--Lindquist coordinates, therefore a coordinate transformation is required to convert the relevant quantities to their corresponding Kerr--Schild versions.
For the density, $S^0$ turns out to be the same in Boyer--Lindquist and Kerr--Schild coordinates due to Equation~\eqref{eq:bltoksVector}, therefore we can obtain the mass and charge densities in Kerr--Schild coordinates as
\begin{equation}
    \rho_{m,s}^{\rm (KS)} \equiv m_s\alpha^{\rm (KS)} S^0_s,\quad \rho_s^{\rm (KS)} \equiv q_s \alpha^{\rm (KS)} S^0_s.
\end{equation}
When initializing particles in the simulations, we inject a fixed number of particles per cell (ppc) in the torus where $\rho_{m,s}^{\rm (KS)} > 0$, and assign each particle a weight proportional to the mass density $\rho_{m,s}^{\rm (KS)}$. To ensure that the initial charge and current density is zero, we inject electrons and positrons in pairs with identical location and momenta.
For particle momenta, we first sample the Boyer--Lindquist momentum $p_{\mu}^{\rm (BL)}$ according to Section~\ref{subsec:momentum_sampling}, then transform to Kerr--Schild coordinates following Equation~\eqref{eq:bltoksCovector}. 
The left column of Figure \ref{fig:density} shows the initial total mass density for a few different realizations of the collisionless torus.
It can be seen that with fixed $r_0$ or $L_0$, a decreasing $r_{\rm in}$, or equivalently, an increasing $E_{\max}$, leads to a larger torus. Note that the density peak is not necessarily located at $r_0$; it is usually at $r_{\rm peak}<r_0$.

We perform a series of tests for the equilibrium torus. The simulations do not include any initial electric or magnetic field, but we do evolve all 6 components of the electromagnetic field.
The computational domain extends from just inside the event horizon to 
\( r \approx 60\,r_g \), well beyond the outer boundary of the torus.
The grid is logarithmically spaced in the radial direction and uniform in the $\theta$ direction. We use a numerical resolution of \( N_r \times N_\theta = 1024 \times 1024 \) and initialize 20 particles per cell.
To properly capture the kinetic physics, we need to resolve the plasma skin depth $\lambda_p=c/\omega_p=\sqrt{m^2 c^2/(4\pi\rho_me^2)}$ and the Debye length $\lambda_D=\sqrt{k_BT_{\rm eff}m/(4\pi \rho_m e^2)}$, where $T_{\rm eff}$ is the effective temperature of the plasma in its rest frame, to be defined in Section~\ref{sec:fluid}. Since our effective temperature is typically non-relativistic, $k_BT_{\rm eff}/mc^2\ll1$, the smallest scale we need to resolve is the Debye length. If the Debye length were not resolved, the plasma would be numerically heated up until the Debye length gets above the grid scale, modifying the initial distribution and generally inflating the torus. In our simulations, we have chosen a density such that at the density peak, the Debye length is $0.1r_g$ and the plasma skin depth is $\sqrt{10}\,r_g$,
well resolved by our grid. This resolution is sufficient to resolve the relevant length scales throughout the torus, despite the fact that the position where the Debye length is least resolved does not coincide exactly with the density peak, and depends on the radial extent of our grid.

We evolve the system for several hundred dynamical times. The final densities for the three cases are shown in the right column of Figure \ref{fig:density}. There is no noticeable evolution in the bulk of the torus; the appearance of noise near the edge of the torus is due to the low density and relatively high particle shot-noise. In particular, there is no inflation of the torus, confirming that no significant numerical heating has occurred. We thus conclude that the initial condition is indeed a steady state.

\section{Fluid Properties}\label{sec:fluid}
We would like to now quantify the fluid-level properties of the plasma in the torus, including its bulk motion, effective temperature, and pressure tensor.

\subsection{Bulk Motion}\label{sec:bulk}

\begin{figure}[ht]
    \centering
    \plotone{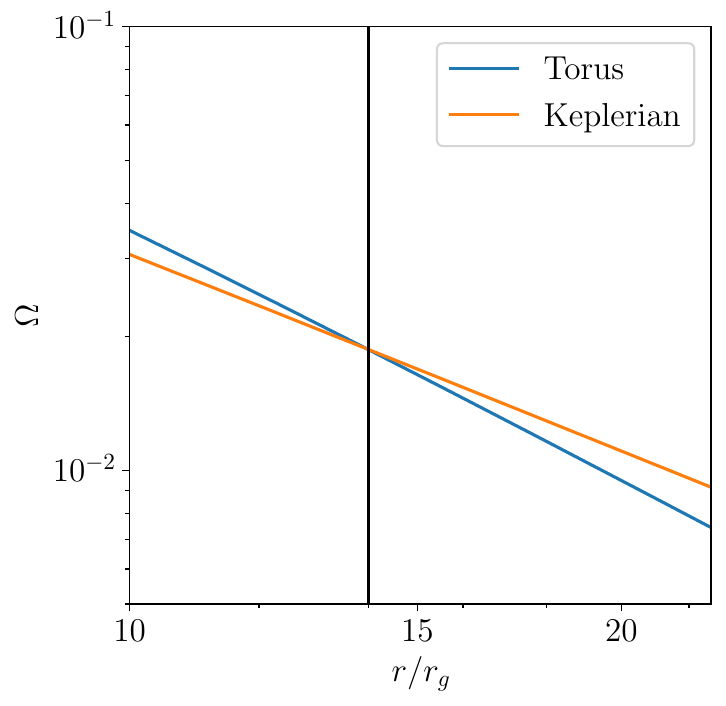}
    \caption{
    Angular velocity on the equatorial plane as a function of radius. The blue line shows the angular velocity of the collisionless torus, defined as $\Omega = S^{\phi} / S^0$, while the orange line shows the angular velocity of stable circular geodesic orbits, $\Omega_C = \left( a + r^{3/2}/\sqrt{M} \right)^{-1}$. The vertical line marks the location of $r_0 = 14$, 
    where the lower bound of the particle energy reaches its minimum.
    The plotted range spans the radial extent of the torus from $r_{\rm in}$ to $r_{\rm out}$.
    }
    \label{fig:angular_velocity}
\end{figure}

The bulk velocity of the fluid is defined as the normalized number-flux four-vector:
\begin{align}
    u^\mu = \frac{S^\mu}{n},
\end{align}
where $n\equiv\left(-S^\mu S_\mu\right)^{1/2}$ is the proper number density. This ensures the proper normalization \( u^\mu u_\mu = -1 \), consistent with the requirement for a timelike four-velocity in general relativity. The angular velocity of the fluid, defined as the rate of change of azimuthal angle with respect to coordinate time, is given by:
\begin{align}\label{eq:angular_velocity}
    \Omega = \frac{d\phi}{dt} = \frac{u^\phi}{u^t} = \frac{S^\phi}{S^t}.
\end{align}
 $\Omega$ is the same in both the Boyer--Lindquist coordinates and the Kerr--Schild coordinates. Figure \ref{fig:angular_velocity} shows the fluid angular velocity $\Omega$ in the equatorial plane. It can be seen that $\Omega$ is super-Keplerian at small radii but sub-Keplerian at large radii. This is similar to the Fishbone-Moncrief torus~\citep{1976ApJ...207..962F}. Due to the differential rotation, we expect this torus to be an excellent basis for studying the magnetorotational instability (MRI) in the accretion flow.

\subsection{Effective Temperature}\label{sec:temp}
The GRPIC framework allows us to track individual particle trajectories and compute thermodynamic quantities from first principles. Since temperature is only well defined in the fluid rest frame (FRF), we first transform particle velocities accordingly.
It is most convenient to construct the FRF tetrad from the Boyer--Lindquist coordinates, as we show in Appendix \ref{sec:frames}. Therefore, to obtain particle velocities in the FRF, we first transform their Kerr--Schild 4-velocity, $v_\mu^{\rm (KS)}$, to Boyer--Lindquist coordinates, $v_\mu^{\rm(BL)}$, then apply the transformation \eqref{eq:FRFtransform} to get the FRF orthonormal 4-velocity $v_{\hat{\mu}}$.
The effective temperature is defined as 
\begin{align}\label{eq:Teff}
    T_{\mathrm{eff}}=\frac{1}{2}m\left<v_{\mathrm{FRF}}^2\right>
\end{align}
where $v_{\mathrm{FRF}} = \left(v_{\hat{i}} v^{\hat{i}}\right)^{1/2}$ is the magnitude of the spatial velocity of the particle species in the FRF. This definition is motivated by the fact that, for a two-dimensional Maxwellian distribution, the temperature corresponds to $m\langle v^2 \rangle/2$. Although our distributions may be truncated, $T_{\mathrm{eff}}$ provides a useful effective measure of the local kinetic energy in the fluid rest frame. 
\begin{figure*}
    \centering    \plotone{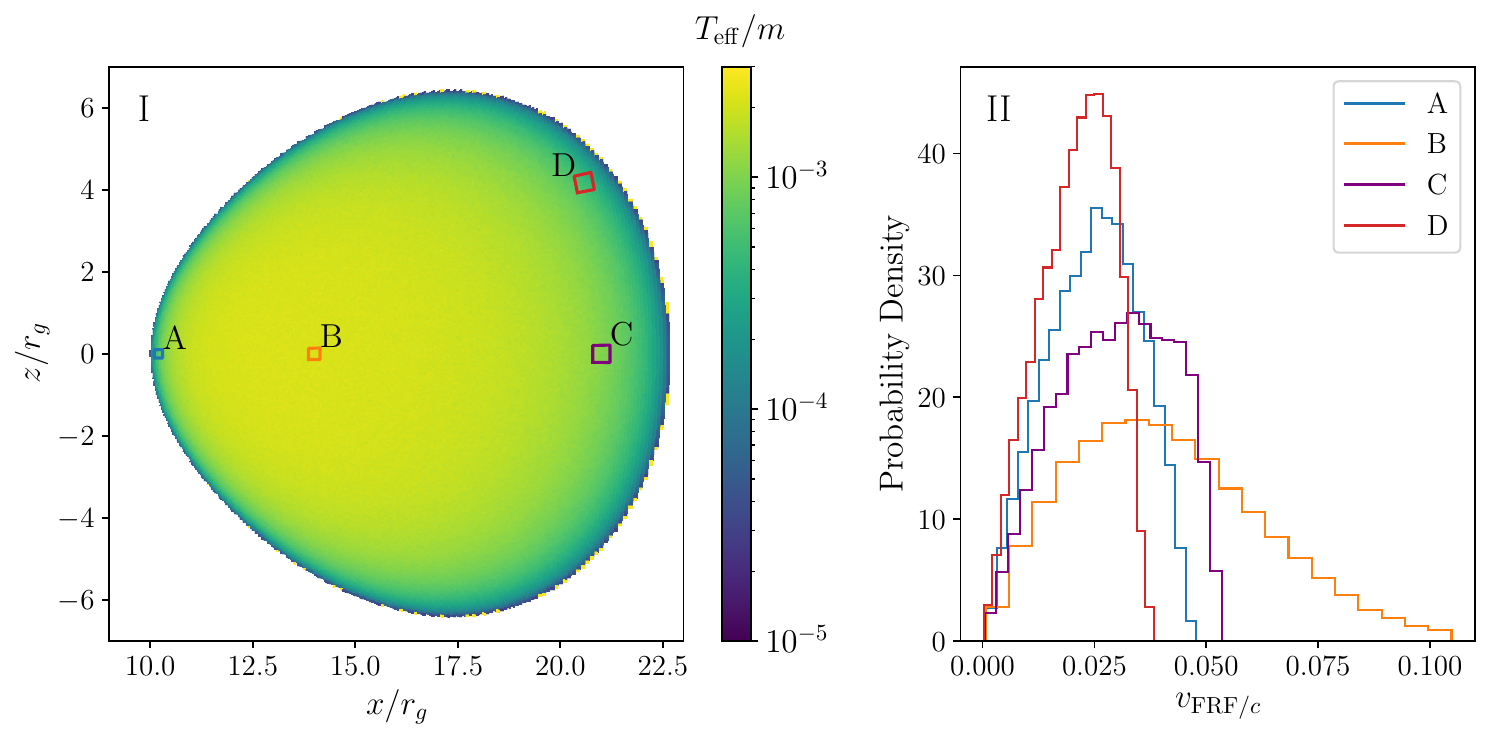}
    \caption{
    \textbf{Panel I:} Cross section of the effective temperature, defined as the average squared particle velocity in the fluid rest frame, see Equation~\eqref{eq:Teff}. The equilibrium corresponds to the middle row of Figure \ref{fig:density}. Four labeled regions (A–D) are outlined, each containing approximately $3\times10^4$ particles. 
    The box shapes follow the local grid structure to preserve consistent particle count. 
    \textbf{Panel II:} Velocity distribution functions $P(v_{\mathrm{FRF}})$ for particles within each region, plotted as probability density (normalized to unity) versus fluid-frame speed $v_{\mathrm{FRF}}$. Each distribution resembles a truncated Maxwellian, with truncation occurring at lower velocities for regions closer to the outer boundary. This illustrates how the energy cutoff in the distribution function varies with position in the torus.
    }
    \label{fig:temp}
\end{figure*}
Panel I of Figure \ref{fig:temp} shows how $T_{\mathrm{eff}}$ changes over the torus, and panel II shows the probability density of $v_{\mathrm{FRF}}$ at different places on the torus. 
For this calculation we used a special run with $1000$ particles per cell to ensure excellent statistics, and the physical parameters are identical to the second row of Figure~\ref{fig:density} with $r_{\rm in} = 10\,r_g$.
It can be seen that near the point $(r=r_0,\theta=\pi/2)$ where $E_{\max}-E_{\min}$ is largest and $T<E_{\max}-E_{\min}$, the distribution is more or less Maxwellian. Toward the edge of the torus where $E_{\min}$ approaches $E_{\max}$, the distribution of $v_{\rm FRF}$ is progressively truncated from the higher end, and the effective temperature goes to zero as expected.

\subsection{Pressure Tensor}\label{sec:pressure}
The stress-energy tensor is computed from the second moment of the distribution function \eqref{equ:distribution}
\begin{align}\label{eq:stress_tensor_integral}
    T_{\mu\nu}=\int p_\mu p_\nu f \frac{d^3p}{p^0 \sqrt{-g}}. 
\end{align}

This expression can be evaluated analytically for the equilibrium distribution, though the resulting form is not compact. The symmetries of the problem simplify the structure of \( T_{\mu\nu} \). In Boyer--Lindquist coordinates, we find:
\begin{itemize}
    \item Constant angular momentum implies \( T_{0\mu} = L_0 S_\mu \),
    \item The off-diagonal components \( T_{r\mu} \) and \( T_{\theta\mu} \) vanish,
    \item The only nonzero mixed components are those involving \( t \) and \( \phi \).
\end{itemize}
Thus, the stress-energy tensor takes the form (the full analytic expression is included in Appendix \ref{sec:stress_integrals}):
\begin{align}
    T_{\mu\nu} =
    \begin{bmatrix}
       T_{00} & 0 & 0 & L_0 S_0 \\
       0 & T_{rr} & 0 & 0 \\
       0 & 0 & T_{\theta\theta} & 0 \\
       L_0 S_0 & 0 & 0 & L_0 S_\phi
    \end{bmatrix}.
\end{align}
To evaluate the fluid pressure, we again need to transform into the local FRF. This transformation can be done analytically, but the expressions are quite lengthy so we only evaluate the stress-energy tensor in the FRF numerically at a sample point as a demonstration. For a torus with $r_0 = 14\,r_g$, $r_{\rm in} = 10\,r_g$, $a = 0.999$, and $T = 10^{-3}m$, the numerical value of $T_{\mu\nu}$ for a single species in the FRF at $(r = r_0, \theta = \pi/2)$ is:
\begin{equation}
    \frac{T_{\mu\nu}^\mathrm{FRF}}{m f_0} \approx
    \begin{bmatrix}
        6.1\times 10^{-2} & 0 & 0 & 1.7\times 10^{-8} \\
        0 & 6.6\times 10^{-5} & 0 & 0 \\
        0 & 0 & 6.6\times 10^{-5} & 0 \\
        1.7\times 10^{-8} & 0 & 0 & 4.6\times 10^{-9}
    \end{bmatrix}.
\end{equation}
It can be seen that the pressures in the $r$ and $\theta$ directions are identical, while the pressure along $\phi$ is effectively zero: there is a strong pressure anisotropy. This is not a coincidence, but simply a reflection of our choice of single angular momentum for all particles $L_z = L_0$. On the other hand, since both $p_r$ and $p_\theta$ are determined by the same Maxwell-Juttner-like distribution (Section~\ref{sec:implementation}), they naturally have the same temperature. In order to allow for pressure anisotropy in the $r$--$\theta$ plane, a distribution over the Carter's constant $Q$ can be potentially introduced in the distribution function defined via Equation~\eqref{equ:distribution}.

Given the large pressure anisotropy, the plasma is prone to the Weibel instability~\citep[see e.g.][]{1973ppp..book.....K}. However, in our 2D simulations, axisymmetry suppresses the unstable modes since their wave vectors are along the colder direction. At the same time, differential rotation may also introduce large-scale phase mixing which has been demonstrated to spontaneously produce seed magnetic fields~\citep{2022PNAS..11919831Z}. Here we are again protected from this type of instability since the shearing flow direction is in our symmetry direction. In a full 3D simulation, we expect that Weibel modes will grow in this setup and spontaneously generate magnetic fields, which will eventually destabilize the torus.

\section{Discussion} \label{sec:discussion}

In this paper, we have constructed a fully analytic equilibrium torus solution with finite angular momentum in the collisionless regime around a Kerr black hole. We realized this torus solution in a 2D axisymmetric GRPIC code and verified that it is indeed stable over hundreds of dynamical times under zero electromagnetic field. By adding a seed magnetic field, this initial condition will become unstable to the magnetorotational instability (MRI), shedding angular momentum and start the accretion process. Therefore, this solution can serve as an ideal initial condition for studying collisionless accretion in a GRPIC framework, allowing the controlled development of MRI and angular momentum transfer, similar to the Fishbone-Moncrief torus widely used in GRMHD simulations. A host of open problems regarding this type of collisionless accretion flow around low-luminosity AGN can now be addressed.

The equilibrium torus distribution we constructed in Equation~\eqref{equ:distribution} has a delta function for $L_z$, meaning that all particles have the same angular momentum. As a result, there is a strong initial pressure anisotropy in the fluid rest frame where $T_\phi$ (the effective temperature of the $\phi$-component motion) is much smaller than $T_r$ and $T_\theta$. It is straightforward to extend the construction in this paper to a distribution function that has a spread in $L_z$, as any function of $L_z$ can replace the delta function while keeping the stationary property of the solution. However, for a general function of $L_z$ we are unable to find simple analytic integrals for $S^0$ and the momentum distribution CDF. Both the density integral and momentum sampling may require Monte Carlo techniques, therefore we omit such a construction in this paper for simplicity.

The strong pressure anisotropy in the initial condition leads to the development of the Weibel instability. 2D axisymmetric simulations suppress this instability, but the instability develops in full 3D setups. If this feature is undesirable, one should be able to finetune the profile of $L_z$ such that the $\phi$ temperature in the fluid rest frame matches the temperature in the other two directions. On the other hand, if sufficient seed magnetic field is put in initially, then it can also suppress the growth of the Weibel instability, allowing the system to develop MRI self-consistently. 

We have briefly experimented with adding a weak magnetic field to the initial condition in a full PIC simulation. When the magnetic field is sufficiently weak (e.g.\ gyroradius much larger than the size of the torus), the torus remains stable for a long time, regardless of the field configuration. With a larger magnetic field such that the plasma is magnetized, gyration tends to equilibrate the pressure along the two directions perpendicular to the magnetic field. As a result, the strong initial pressure anisotropy in our equilibrium torus quickly develops into an anisotropy with respect to the magnetic field direction, $p_\perp < p_\parallel$. In the center of the torus where $\beta \gg 1$, this leads to the development of the firehose instability~\citep[see e.g.][]{2014PhRvL.112t5003K} that brings the plasma closer to isotropy. Long term evolution of this scenario invariably leads to the development of MRI which initiates the accretion of the torus onto the black hole. We will explore the physics of collisionless accretion in a future work.

Since the collisionless torus has a finite size, the initial condition is entirely vacuum outside of the torus. This would have posed problems for GRMHD codes since they would fail at zero mass density, but GRPIC can tolerate vacuum with no problem. For direct comparison with GRMHD simulations however, some type of charge injection may be needed, either through some artificial injection scheme~\citep{2019PhRvL.122c5101P,2023PhRvL.130k5201G}, or via more elaborate self-consistent pair production schemes~\citep{2020PhRvL.124n5101C,2025ApJ...985..159Y,2025PhRvL.135a5201V}. In particular, such charge injection could be crucial in the funnel region of the jet to maintain the Blandford-Znajek process and carry the necessary Poynting flux.

\begin{acknowledgments}
We thank Beno\^it Cerutti, Charles Gammie, John Mehlhaff, Dmitri Uzdensky, Vladimir Zhdankin, and Muni Zhou for helpful discussions. We particularly thank John Mehlhaff for very useful comments on the manuscript.
AC and YY acknowledge support from NSF grants DMS-2235457 and
AST-2308111. AC acknowledges additional support from NASA grant
80NSSC24K1095. YY also acknowledges support from the Multimessenger Plasma
Physics Center (MPPC), NSF grant PHY-2206608, and support from the Simons
Foundation (MP-SCMPS-00001470). We thank the hospitality of the Kavli Institute for Theoretical Physics (KITP), where this manuscript is finished. This research was supported in part by grant NSF PHY-2309135 to the KITP.
\end{acknowledgments}

\appendix

\section{Metric}\label{sec:metric}
The Kerr spacetime can be parameterized in Boyer--Lindquist coordinates
\begin{align} \label{blmetric}
    ds^2=-\left(1-\frac{2Mr}{\Sigma}\right)dt^2
    +\frac{\Sigma}{\Delta}dr^2+\Sigma d\theta^2
    +\frac{A}{\Sigma}\sin^2\theta d\phi^2
    -\frac{4Mar \sin^2\theta }{\Sigma}dtd\phi,
\end{align}
where $\Delta=r^2-2Mr+a^2$, $\Sigma=r^2+a^2 \cos^2\theta$ and $A=(r^2+a^2)^2-a^2\sin^2\theta \Delta$.
Using the transformation given by e.g.,\ ~\citet{2004MNRAS.350..427K}, we can convert a vector $V^\mu$ from Boyer--Lindquist into Kerr--Schild ${V^\mu}'$
\begin{align}\label{eq:bltoksVector}
    {V^0_{\rm (KS)}}=V_{\rm (BL)}^0+\frac{2Mr}{\Delta}V^r_{\rm (BL)},
    &&{V^r_{\rm (KS)}}=V_{\rm (BL)}^r,
    &&{V^\theta_{\rm (KS)}}=V_{\rm (BL)}^\theta,
    &&{V^\phi_{\rm (KS)}}=V_{\rm (BL)}^\phi+\frac{a}{\Delta} V^r_{\rm (BL)},
\end{align}
and for a covector $V_{\mu}$
\begin{align} \label{eq:bltoksCovector}
    {V^{\rm (KS)}_0}=V^{\rm (BL)}_0,
    &&V^{\rm(KS)}_r=V^{\rm (BL)}_r-\frac{a V^{\rm (BL)}_\phi+2Mr V^{\rm (BL)}_0}{\Delta}, &&
    &V^{\rm(KS)}_\theta=V^{\rm (BL)}_\theta, &&
    &V^{\rm(KS)}_\phi=V^{\rm (BL)}_\phi.
\end{align}
We can write the Kerr spacetime in terms of Kerr--Schild coordinates
\begin{align} \label{ksmetric}
    ds^2=&-\left(1-\frac{2Mr}{\Sigma}\right)dt^2
    +\left(1+\frac{2Mr}{\Sigma}\right)dr^2+\Sigma d\theta^2
    +\frac{A}{\Sigma}\sin^2\theta d\phi^2
    \\&+\frac{4Mr}{\Sigma}dtdr
    -2 a \left(1+\frac{2Mr}{\Sigma}\right)\sin^2\theta drd\phi
    -\frac{4Mar \sin^2\theta }{\Sigma}dtd\phi\,.
\end{align}

\section{3+1 Splitting}\label{sec:splitting}
In the 3+1 formalism, spacetime splits into 3-dimensional spacelike hypersurfaces $\Sigma_t$. Each slice is simultaneous, having constant time $t$, as measured by a local fiducial observer (FIDO) at rest with respect to the slice. The four-velocity $n_\mu$ of the FIDO is normal to $\Sigma_t$ and is given by
 \begin{align}
    n_{\mu} &=(-\alpha, 0^i),\\
    n^{\mu} &=(1,-\beta^i)/\alpha\,.
\end{align}
 Where $\alpha$ is the the lapse function which measures the proper time between slices. The metric $g_{\mu\nu}$ and its inverse can be written in terms of $\gamma_{ij}$ the spatial metric of $\Sigma_t$, the lapse function $\alpha$ and the shift vector $\beta^i$ which connects points with the same coordinates between slices
\begin{align}
    g_{\mu\nu} =
    \begin{bmatrix}
        \beta^2-\alpha^2 & \beta^j \\
        \beta^i & \gamma_{ij}
    \end{bmatrix}, &&
    g^{\mu\nu} =
    \frac{1}{\alpha^2}
    \begin{bmatrix}
        -1 & \beta^j \\
        \beta^i & \alpha^2\gamma^{ij} -\beta^i\beta^j
    \end{bmatrix}.
\end{align}
Noting that the spatial metric $\gamma_{ij}$ is used to raise and lower spatial vectors e.g. $\beta_i= \gamma_{ij}\beta^j$. We assume the metric is stationary $\partial_t g=0$ which implies $\alpha$, $\beta^i$ and $\gamma_{\mu\nu}$ are also time independent. Explicitly these the Boyer--Lindquist components are (spatial components are ordered by $r$, $\theta$, $\phi$):
\begin{align}\label{BL3+1}
    \alpha=\sqrt{\frac{\Delta\Sigma}{A}}\,,&&
    \beta^i=\left(0,0,-\frac{2Mar}{A}\right), &&
     \gamma^{ij}=
     \begin{bmatrix}
        \displaystyle\frac{\Delta}{\Sigma} &0&0\\
        0 & \displaystyle\frac{1}{\Sigma} &0\\
        0 & 0 & \displaystyle\frac{\Sigma}{A\sin^2\theta}
    \end{bmatrix},
\end{align}
and in Kerr--Schild
\begin{align}\label{KS3+1}
    \alpha=\sqrt{\frac{\Sigma}{\Sigma+2Mr}}\,,&&
    \beta^i =\left(\frac{2Mr}{\Sigma+2Mr},0,0\right), &&
     \gamma^{ij}=
     \begin{bmatrix}
       \displaystyle\frac{A}{\Sigma(\Sigma+2Mr)} & 0 & \displaystyle\frac{a}{\Sigma}\\
        0 & \displaystyle\frac{1}{\Sigma} & 0\\
        \displaystyle\frac{a}{\Sigma} & 0 & \displaystyle\frac{1}{\Sigma\sin^2\theta}
    \end{bmatrix}.
\end{align}
Let the 4-momentum of a particle be $p_\mu=\partial x_\mu/\partial\lambda$. The spatial momentum is given by $p_i$. We can easily write the contravariant components of the 4-momentum:
\begin{align}
    p^\mu &=g^{\mu\nu} p_\nu \\ 
    p^0 &= g^{tt} p_0+g^{0i} p_i= \frac{1}{\alpha^2}\left(\beta^i p_i-p_0\right)\\
    p^i &= g^{it} p_0+g^{ij} p_j= \frac{1}{\alpha^2}\left(\beta^i (p_0-\beta^j p_j) +\alpha^2\gamma^{ij}p_j\right)\notag \\
    &=\gamma^{ij} p_j-p^0\beta^i\,.
\end{align}
Note that $p^i$ is distinct from the raised spatial velocity $\gamma^{ij}p_j$. We generally avoid using the symbol $p^i$ due to this ambiguity.
We can also use the normalization of $p_\mu$
\begin{align}
\notag -m^2 &=g^{\mu\nu}p_\mu p_\nu\\
\notag &=g^{tt}p_0^2+2g^{it}p_i p_0+g^{ij}p_i p_j\\
&=\frac{1}{\alpha^2} \left(-p_0^2+2\beta^{i}p_i p_0+\left(\alpha^2\gamma^{ij}-\beta^i\beta^j\right)p_i p_j\right).
\end{align}
The quadratic formula gives a second definition of $p^0$ and $p_0$:
\begin{align}
\label{p_0}
p_0 &=\beta^i p_i- \alpha \sqrt{\gamma^{ij}p_i p_j+m^2}\,,\\
\label{p^0}
p^0 &= \frac{1}{\alpha}\sqrt{\gamma^{ij}p_i p_j+m^2}\,.
\end{align}
The ambiguity of $\pm$ was resolved by choosing $p_\mu$ to be future directed.
\section{Separation of Variables} \label{sec:separation}
Following \citep{1968PhRv..174.1559C}, the Hamilton-Jacobi equation is
\begin{align} \label{HamiltonJacobi}
    -\frac{\partial S}{\partial \lambda}=\mathcal{H}\left(x^\mu,\frac{\partial S}{\partial x^\mu}\right)=\frac{1}{2}g^{\mu\nu}\frac{\partial S}{\partial x^\mu}\frac{\partial S}{\partial x^\nu}\,.
\end{align}
Here the substitution $p_\mu=\frac{\partial S}{\partial x^\mu}$ was made. Carter introduces the general solution for the Jacobi action assuming the equations are separable
\begin{align} \label{Action}
    S=\frac{1}{2}m^2 \lambda-E t+S_r(r)+S_\theta(\theta)+L_z \phi\,.
\end{align}
Substituting this into~\eqref{HamiltonJacobi}
\begin{align} \label{HamiltonJacobiAction}
    \notag 0&=m^2+g^{tt}E^2+g^{rr}\left(\frac{\partial S_r}{\partial r}\right)^2+g^{\theta\theta}\left(\frac{\partial S_\theta}{\partial \theta}\right)^2+g^{\phi\phi}L_z^2-2g^{t\phi}EL_z\\ 
      &=\Sigma m^2-\frac{A}{\Delta}E^2+\Delta\left(\frac{\partial S_r}{\partial r}\right)^2+\left(\frac{\partial S_\theta}{\partial \theta}\right)^2+\frac{\Delta-a^2\sin^2\theta}{\Delta \sin^2\theta}L_z^2+\frac{4Mar}{\Delta}EL_z.
\end{align}
Separating this equation into two parts one depending on r and the other on $\theta$ we have
\begin{gather}\label{HamiltonJacobiSeparation1}
    \notag-r^2m^2+\frac{(a^2-r^2)^2}{\Delta}E^2-\Delta\left(\frac{\partial S_r}{\partial r}\right)^2+\frac{a^2}{\Delta}L_z^2-\frac{4Mar}{\Delta}EL_z=\\a^2 m^2 \cos^2\theta+a^2E^2\sin^2\theta +\left(\frac{\partial S_\theta}{\partial \theta}\right)^2+\frac{L_z^2}{\sin^2\theta}\,.
\end{gather}
Subtracting $2aEL_z$ from both sides
\begin{gather}\label{HamiltonJacobiSeparation2}
        -\Delta\left(\frac{\partial S_r}{\partial r}\right)^2+\frac{\left((r^2-a^2)E-a L_z\right)}{\Delta}-r^2m^2=\left(\frac{\partial S_\theta}{\partial \theta}\right)^2+a^2 m^2 \cos^2\theta+\left(aE\sin\theta-\frac{L_z}{\sin\theta}\right)^2
\end{gather}
The two sides must be equal to a constant $C$ because they are independent of each other. This is related to Carter's constant $Q=C-(L_z-aE)^2$. $C$ is always nonnegative whereas $Q$ can be negative. Combining this with the right side of \eqref{HamiltonJacobiSeparation2}
\begin{gather}\label{CarterConstant}
    Q=p_\theta^2+\left(a^2(m^2-E^2)+\frac{L_z^2}{\sin^2\theta}\right)\cos^2\theta\,.
\end{gather}
We can substitute back into \eqref{Action} to get 
\begin{gather}
    S=-\frac{1}{2}m^2 \lambda-E t\pm\int \frac{\sqrt{R}}{\Delta}dr\pm \int \sqrt{\Theta}d\theta+L_z \phi\,,
\end{gather}
where
\begin{gather}
    R=\left(E(r^2+a^2)-aL_z\right)^2-\Delta\left(r^2+(L_z-aE)^2+Q\right)\label{eq:radial_function},\\
    \Theta=Q-\left(a^2(m^2-E^2)+\frac{L_z^2}{\sin^2\theta}\right)\cos^2\theta\label{eq:polar_function}.
\end{gather}
Differentiating Equation \eqref{Action}, as in $p_\mu=\frac{\partial S}{\partial x^\mu}$, we derive \eqref{equ:momentum}.
\section{Circular Orbits}
\label{sec:circular_orbits}
Circular orbits correspond to trajectories that remain at constant radius. This requires the radial motion to be at a turning point with zero radial velocity and zero radial acceleration, which imposes the conditions:
\begin{align}
    R(r) = 0\,, \qquad \frac{dR}{dr} = 0\,.
\end{align}
Solving these two equations simultaneously yields the specific angular momentum and energy for circular equatorial orbits at radius $r_0$ which give \eqref{eq:circular_amom} and \eqref{eq:circular_energy} respectively. As in \eqref{eq:angular_velocity}, the angular velocity $\Omega_C$ of a particle in circular motion is given by
\begin{align}
    \Omega_C = \frac{d\phi}{dt} = \frac{u^\phi}{u^t}\,.
\end{align}
Using the normalization condition $u^\mu u_\mu = -1$ and substituting the expressions for $E$ and $L_z$ back into the geodesic equations, one finds
\begin{align}
    \Omega_C = \frac{1}{\left( a + r^{3/2}/\sqrt{M} \right)}\,.
\end{align}

\section{Tetrad of the fluid rest frame}\label{sec:frames}
To evaluate the fluid properties of the torus, we must compute quantities in the rest frame of the bulk flow, the fluid rest frame (FRF). Here we construct the transformation from Boyer--Lindquist coordinates directly into the FRF. In the FRF, the bulk four velocity of the fluid must be $u^{\hat\mu}=(1,0,0,0)$. Thus the transformation $e^{\mu}_{\hspace{0.3em}{\hat{\nu}}}$ must take $u^{\hat\mu}\to u^\mu$. So
\begin{align}
u^\mu=e^{\mu}_{\hspace{0.3em}\hat{\nu}}u^{\hat\nu}=e^{\mu}_{\hspace{0.3em}\hat{0}} \,.
\end{align}
This constrains the zeroth basis vector to $e_{\hat 0}=u^0\partial_t+u^\phi\partial_\phi$. Since Boyer--Lindquist coordinates are diagonal for $r,\theta$, we choose $e_{\hat r}=g_{rr}^{-1/2}\partial_r$ and $e_{\hat \theta}=g_{\theta\theta}^{-1/2}\partial_\theta$. This structure maintains the direction of the $r$ and $\theta$ components. We choose the final basis vector to be orthogonal to the rest. This has the form $e_{\hat \phi}=c_1\partial_t+c_2\partial_\phi$. The orthogonality ensures that
\begin{align}
    0&=e_{\hat 0}\cdot e_{\hat \phi}\notag\\
    &=g_{\mu\nu}e^{\mu}_{\hat 0} e^{\nu}_{\hat \phi}\notag\\
    &=g_{00}u^0c_1+g_{0\phi}u^0c_2+g_{\phi0} c_1 u^\phi+g_{\phi\phi}u^\phi c_2\notag\\
    &=u_0c_1+u_\phi c_2
\end{align}
and normalization requires that
\begin{align}
    1=e_{\hat \phi}\cdot e_{\hat \phi}=g_{00}c_1^2+2g_{0\phi}c_1c_2+g_{\phi\phi}c_2^2\,.
\end{align}
With these constraints, we can derive $c_1$ and $c_2$. The same approach can be used to derive the inverse transformation. In summary, these two transformations are 
\begin{align}\label{eq:FRFtransform}
    e^{\mu}_{\hspace{0.3em}{\hat{\nu}}}=
    \begin{bmatrix}
    u^{0} & 0 & 0 & c_{1} \\
    0 & g_{rr}^{-1/2} & 0 & 0 \\
    0 & 0 & g_{\theta\theta}^{-1/2} & 0 \\
    u^{\phi} & 0 & 0 & c_{2}
    \end{bmatrix}, &&
    e^{\hat\mu}_{\hspace{0.3em}{\nu}}=
    \begin{bmatrix}
    u_{0} & 0 & 0 & u_{\phi} \\
    0 & \sqrt{g_{rr}} & 0 & 0 \\
    0 & 0 & \sqrt{g_{\theta\theta}} & 0 \\
    d_{1} & 0 & 0 & d_{2}
    \end{bmatrix},
\end{align}
where 
\begin{align}
    c_{1} &= \left[\,g_{00} - 2\,g_{0\phi}\frac{u_0}{u_{\phi}} + g_{\phi\phi}\left(\frac{u_{0}}{u_{\phi}}\right)^{2}\,\right]^{-1/2},
     && 
    c_{2} = -\frac{u_{0}}{u_{\phi}}\,c_{1}\,,
\end{align}
and
\begin{align}
d_{1} &= \left[g^{00} - 2\,g^{0\phi} \frac{u^{0}}{u^{\phi}} + g^{\phi\phi} \left(\frac{u^{0}}{u^{\phi}}\right)^{2}\,\right]^{-1/2},
&& d_{2} = -\frac{u^{0}}{u^{\phi}}\,d_{1}\,.
\end{align}
\section{Stress Tensor}\label{sec:stress_integrals}
In Section~\ref{sec:pressure}, we described that the components of the stress energy tensor can be integrated analytically in Boyer--Lindquist coordinates. To evaluate these integrals, we expand \eqref{eq:stress_tensor_integral} using the coordinate transform \eqref{equ:transformation} as
\begin{align}
        T_{\mu\nu} =\frac{\alpha C f_0\exp\left(\displaystyle\frac{E_{\max}+\beta^{\phi}L_0}{T}\right)}{\sqrt{\Delta \sin^2\theta}}\int \frac{p_\mu p_\nu\hat p }{\sqrt{\hat p^2+1}} \exp\left(-\frac{\sqrt{\hat p^2+1} }{\hat{T}}\right)d\hat{p}\,d\psi\,.
\end{align}
There are three nonzero components that are not just scalings of $S_\mu$, they are:
\begin{align}
T_{00} &= \frac{ 2 \pi T f_0 }
{\sqrt{ \Delta \, \sin^2\theta }}\left(\left(E+T\right)^2+T^2 \right) \left.\exp\left(\frac{E_{\max}-E }{T}\right)\right|_{E_{\max}}^{E_{\min}}\label{eq:T00}\,, \\
T_{rr} &= \frac{ g_{rr}\pi T f_0 }
{ \alpha^2  \sqrt{\Delta\sin^2\theta} } \left(\left(E + T + L_0 \beta^{\phi} \right)^2+ T^2 - \alpha^2 C^2  \right)\left.\exp\left(\frac{E_{\max}-E }{T}\right)\right|_{E_{\max}}^{E_{\min}}\label{eq:Trr}\,, \\
T_{\theta\theta} &= \frac{ g_{\theta\theta}\pi Tf_0 }
{ \alpha^2\sqrt{\Delta\sin^2\theta} } \left(\left(E + T + L_0 \beta^{\phi} \right)^2+ T^2 -\alpha^2 C^2  \right) \left.\exp\left(\frac{E_{\max}-E }{T}\right)\right|_{E_{\max}}^{E_{\min}}\label{eq:Tthetatheta}\,.
\end{align}

\bibliography{references}
\bibliographystyle{aasjournalv7}

%% This command is needed to show the entire author+affiliation list when
%% the collaboration and author truncation commands are used.  It has to
%% go at the end of the manuscript.
%\allauthors

%% Include this line if you are using the \added, \replaced, \deleted
%% commands to see a summary list of all changes at the end of the article.
%\listofchanges

\end{document}